\documentclass[prd,aps,preprint,amsmath,nofootinbib,amssymb,eqsecnum,showkeys,tightenlines]{revtex4-1}

\usepackage{amsmath, latexsym, amssymb, hyperref, graphicx, color, multirow, makecell, diagbox}
\usepackage[table]{xcolor}
\usepackage{tabularx}
\usepackage[T1]{fontenc}
\usepackage[capitalize]{cleveref}
\definecolor{nicered}{rgb}{.7,.1,.1}
\definecolor{nicegreen}{rgb}{.1,.5,.1}
\definecolor{darkblue}{rgb}{0,0,.5}
\hypersetup{colorlinks, citecolor=nicegreen,linkcolor=nicered, urlcolor=darkblue}
\usepackage{multirow}
\usepackage{slashed}
\numberwithin{equation}{section}
\usepackage{verbatim}

\begin{document}
\preprint{}

\title{Comments on the expanded Maxwell's equations for moving charged media system}

\author{Chuang Li$^{1}$}
\email{lichuang@alumni.itp.ac.cn}

\author{Junle Pei$^{2,3}$}
\email{peijunle@mail.itp.ac.cn, corresponding author}

\author{Tianjun Li$^{2,3}$}
\email{tli@mail.itp.ac.cn}

\affiliation{$^1$ College of Mechanical and Electrical Engineering, Wuyi University, Nanping 354300, China}
\affiliation{$^2$CAS Key Laboratory of Theoretical Physics, Institute of Theoretical Physics, Chinese Academy of Sciences, Beijing 100190, China}
\affiliation{$^3$ School of Physical Sciences, University of Chinese Academy of Sciences,
No.~19A Yuquan Road, Beijing 100049, China}

\begin{abstract}

 In the recent work~\cite{Wang:2021p2}, the author proposed the expanded Maxwell's equations for moving charged media system, which seems subtle.
Considering a very short time, we can approximately define the inertial frame of reference. If we assume all the physical quantities are defined in the same reference frame by default, Maxwell's equations for the static media system and moving media system are definitely the traditional Maxwell's equations, which are covariant and consistent with the two fundamental postulates 
of special relativity. We even prove the covariance of Maxwell's equations explicitly by considering the Lorentz transformation under the ${\cal O} (v)$ order approximation and the Galileo approximation, respectively. Therefore, it seems to us that the fields in the expanded Maxwell's equations cannot be in the same reference frame. Defining the fields in the lab and co-moving frames explicitly, we derive the expanded Maxwell's equations for moving media system. Furthermore, we discuss another possible variant of Maxwell's equations, which has an additional coefficient $\alpha$ related to the media. However, it is still subtle from theoretical point of view.

\end{abstract}


\maketitle

\section{Introduction}\label{sec:intro}

Since Maxwell wrote down his great electromagnetism (or classical electrodynamics modernly) equations~\cite{Maxwell:1861}, which unified all the electric and magnetic phenomena, the world has changed hugely because of these equally great experimental proofs and applications of these equations, such as Hertz's experiment and Tesla's alternating current invention, etc. Later, Einstein studied the theory of electrodynamics of moving bodies, and
proposed his great theory of special relativity~\cite{Einstein:1905}. Classical electrodynamics was the first physical theory consistent with special relativity. Right now, the Lorentz covariance of special relativity 
has become a basic requirement of theories for particle physics (or quantum field theory in other words), 
which precisely describe the physical laws of three fundamental interactions in the nature except gravity.
These theories have gotten sufficient verification and comprehensive acceptance~\footnote {Please check the textbooks, such as ~\cite{Jackson:1999, Landau-Lifshitz:ED_for_CM, GuoShuoHong:ED, CaoChangQi:CED, ZhaoKH-ChenXM:EM, QFT_Books} etc.}. 
 
Although the wide recognition, because of the non-trivial four-vector and four-tensor form of Lorentz transformation and the common phenomenon of slow moving media, {\it i.e.}, the media moving speed $v$ is much smaller
 than light speed $c$, there are intermittent works devoting to get a convenient simplified theory of 
Maxwell's equations, such as the works in Refs~\cite{Hertz:1890, Minkowski:1908, Kaufman:1962, Darrigol:1995,  Tai:1964, Costen-Adamson:1965, Rozov:2015, Rozov:2017, Bellac-LevyLeblond:1973, Rousseauxa:2013}. 
 
Recently, the work in Ref~\cite{Wang:2021p2} proposed a new expansion of Maxwell equations for moving charged media system based on the application scenes of triboelectric nanogenerators (TENGs). The expanded equations include new $\vec{v} \cdot \nabla$ terms as compared to the standard Maxwell's equations. Let us present the new equations in Ref~\cite{Wang:2021p2} as follows~\footnote{Please see page 11 in~\cite{Wang:2021p2}.}
\begin{align}
	&\nabla \cdot \vec{D}^{\prime} =\rho_{f} - \nabla \cdot \vec{P}_{s}~,\\
	&\nabla \cdot \vec{B} = 0~,\\
	&\nabla \times \vec{E} = -\left({\frac{\partial }{\partial t}} + \vec{v} \cdot \nabla\right)\vec{B}~, \\
	&\nabla \times \vec{H} =\vec{J}_{f} +\left({\frac{\partial }{\partial t}} + \vec{v} \cdot \nabla\right)\left(\vec{P}_{s} +\vec{D}^{\prime}\right)~,
\end{align} 
where $\vec{P}_{s}$ is the polarization density term owing to electrostatic charges on medium surfaces as produced by effect such as triboelectrification, introduced by the Refs~\cite{Wang:2017p1,Wang:2017p2,Wang:2020p1,Wang:2020p2,Wang:2021p1}, and $\vec{D}^{\prime}$ is the normal electric displacement field.

Although the term $\vec{P}_{s}$ is important in the application of TENGs, it's not relevant with our discussions here. For simplicity, we define the total electric displacement field $\vec{D}\equiv \vec{D}^{\prime} + \vec{P}_{s}$, and 
discuss the following equations as the results of Ref~\cite{Wang:2021p2}
\begin{align}
	&\nabla \cdot \vec{D} =\rho_{f} \label{w1}~,\\
	&\nabla \cdot \vec{B} = 0\label{w2}~,\\
	&\nabla \times \vec{E} = -\left({\frac{\partial }{\partial t}} + \vec{v} \cdot \nabla\right)\vec{B} \label{w3}~,\\
	&\nabla \times \vec{H} =\vec{J}_{f} +\left({\frac{\partial }{\partial t}} + \vec{v} \cdot \nabla\right) \vec{D}~.
	\label{w4}
\end{align}

Eqs.~(\ref{w3}) and (\ref{w4}) are different from the standard Maxwell's equations~\footnote{Please check any standard text book of electrodynamics, such as ~\cite{GuoShuoHong:ED, ZhaoKH-ChenXM:EM, Jackson:1999, Landau-Lifshitz:ED_for_CM}, etc.}, and thus are not Lorentz covariant, {\it i.e.}, not consistent with special relativity. This feature incurs discussions and controversies~\cite{Wang-Yang:2022, WangQing:2022p1, WangQing:2022p2}.

Frankly speaking, many people (especially those with theoretical physics background), including us, do not agree with these expansions at the first sight, according to the standard text books of electrodynamics. By deeply analyzing the text books and related literatures, we find that equations  similar to those of Ref~\cite{Wang:2021p2} in forms can be obtained, but fields in these equations are not in the same reference frame, which seems to be different from the descriptions of Ref~\cite{Wang:2021p2}.

The outline of this paper is as follows.
In Sec.~\ref{sec2}, a concise retrospect of the classical electrodynamics from the very beginning of quantum field theory is given.
In Sec.~\ref{sec3}, we think this topic at a higher level and from some essential principles, and declare our basic standpoint.
In Sec.~\ref{sec4} and \ref{sec5}, we explicitly prove that Maxwell's equations are covariant by considering the Lorentz transformation from the co-moving frame of the media to the lab frame under the ${\cal O} (v)$ order approximation and the Galileo approximation, respectively.
In Sec.~\ref{sec6}, equations similar to those of Ref~\cite{Wang:2021p2} in forms are obtained, but fields in these equations are not in the same reference frame.
In Sec.~\ref{sec7}, using the Lorentz covariance, 
we discuss another possible variant of Maxwell's equations for slow moving media, 
in the sense of special definitions of fields.
Sec.~\ref{sec8} provides a conclusion.

Convention statement: we will preferentially use Natural Units to do the formula derivation for simplification, which means $c=\epsilon_0=\mu_0=1$.
%%%%%%%%

\section{Retrospect and Results of Classical Electrodynamics}\label{sec2}

The standard model (SM) of particle physics based on the quantum field theory \cite{QFT_Books} is one of the cornerstones of modern physics, and has achieved a great success. It includes the quantization extension of Maxwell's theory, denoted as the quantum electrodynamics (QED), and can also be used to describe the weak and strong interactions. 
According to the SM, the Lagrangian of QED is expressed as
\begin{align}
\mathcal{L}_{\text{QED}}=-\frac{1}{4}F_{\mu\nu}F^{\mu\nu}+\bar{\psi}i\gamma^\mu\left(\partial_\mu +iQe A_\mu\right) \psi-m\bar{\psi}\psi~,
\end{align}
where $F^{\mu\nu}$ is a four-tensor defined as
\begin{align}
F_{\mu\nu}&=\partial_\mu A_\nu-\partial_\nu A_\mu=\left[\begin{array}{cccc}
0 & E_{x} & E_{y} & E_{z} \\
-E_{x} & 0 & -B_{z} & B_{y} \\
-E_{y} & B_{z} & 0 & -B_{x} \\
-E_{z} & -B_{y} & B_{x} & 0
\end{array}\right]
\label{F_Tensor}~.
\end{align}
Using Euler-Lagrange equation, we get
\begin{align}
\partial_\mu F^{\mu\nu}=J^\nu \label{maxwell1}
\end{align}
with
\begin{align}
J^\nu=Qe\bar{\psi}\gamma^\nu\psi=(\rho,\vec{J})~.
\end{align}
From the Bianchi identity, we obtain
\begin{align}
\partial_\lambda F_{\mu\nu}+\partial_\mu F_{\nu\lambda}+\partial_\nu F_{\lambda\mu}=0~. \label{maxwell2}
\end{align}
Eq.~(\ref{maxwell1}) gives Maxwell's equations with source
\begin{align}
&\nabla\cdot\vec{E}=\rho~,\\
&\nabla\times\vec{B}=\frac{\partial \vec{E}}{\partial t}+\vec{J}~.
\end{align}
Eq.~(\ref{maxwell2}) gives the source free Maxwell's equations 
\begin{align}
&\nabla\cdot\vec{B}=0~,\\
&\nabla\times\vec{E}=-\frac{\partial \vec{B}}{\partial t}~.
\end{align}
Taking into account the media effects, we can rewrite Maxwell's equations as follows
\begin{align}
&\nabla\cdot\vec{D}=\rho_f~,\label{m1}\\
&\nabla\cdot\vec{B}=0~,\label{m2}\\
&\nabla\times\vec{E}=-\frac{\partial \vec{B}}{\partial t}~,\label{m3}\\
&\nabla\times\vec{H}=\vec{J}_f+\frac{\partial \vec{D}}{\partial t}\label{m4}~,
\end{align} 
with
\begin{align}
&\vec{D}=\vec{E}+\vec{P}~,\\
&\vec{H}=\vec{B}-\vec{M}~,
\end{align}
where
$\vec{P}$ and $\vec{M}$ are the electric and magnetic polarization densities, respectively. It is noted that $\vec{D}$ and $\vec{H}$ also form a tensor in four-dimensional space-time~\cite{Landau-Lifshitz:ED_for_CM:SC}
 \begin{align}
 K_{\mu\nu}&=\left[\begin{array}{cccc}
 0 & D_{x} & D_{y} & D_{z} \\
 -D_{x} & 0 & -H_{z} & H_{y} \\
 -D_{y} & H_{z} & 0 & -H_{x} \\
 -D_{z} & -H_{y} & H_{x} & 0
 \end{array}\right]~.
 \label{K_Tensor}
 \end{align}
 Under a Lorentz boost of $\vec{\beta}$ ($\vec{\beta}$ is along the $z$-axis), any four-tensor transforms as
 \begin{align}
 F^{\prime\mu\nu}&=\Lambda^\mu_\alpha\Lambda^\nu_\beta F^{\alpha\beta}~,
 \end{align}
where 
 \begin{align}
	\Lambda&=\left[\begin{array}{cccc}
		\gamma &   0   & 0      & -\beta\gamma \\
		0      &   1   & 0      & 0 \\
		0      &   0   & 1      & 0 \\
		-\beta\gamma & 0 & 0    & \gamma
	\end{array}\right]~,\\
    \gamma&= \frac {1} {\sqrt{1-\beta^2}}~.
\end{align}
 Demonstrating in three-dimensional vectors, we obtain the Lorentz transformations as
  \begin{align}
  &E^\prime_{\|}=E_{\|}~,\\
  &B^\prime_{\|}=B_{\|}~,\\
   &E^\prime_{\perp}=\gamma\left(\vec{E}+\vec{v}\times\vec{B} \right)_{\perp}~, \\
  &B^\prime_{\perp}=\gamma\left(\vec{B}-\vec{v}\times\vec{E} \right)_{\perp}~,
 \end{align}
 where the subscript of ${\|}$ ($\perp$) characters the component parallel (perpendicular) to $\vec{\beta}$ of the relevant three-dimensional vector.
The transformations of $\vec{D}$ and $\vec{H}$ are similar to $\vec{E}$ and $\vec{B}$ under the same Lorentz boost.
 
 \section{Theoretical Thinking}\label{sec3}
 
 We shall provide the theoretical considerations in this Section.
 First, to study Maxwell's equations, we need to consider the frame of reference. 
 Maxwell's equations in the non-inertial frame are very complicated, and then we do not
 think that the author of Ref. \cite{Wang:2021p2} considered it. However, 
 there might exist acceleration in the co-moving frame, and we should define the inertial frame concretly.
 Considering very short time, {\it i.e.}, $\Delta t$ is very small, we have
 \begin{align}
 \Delta \vec{s} = \vec{v} \Delta t + {1\over 2} \vec{a} \Delta t^2~.~
 \end{align}
 When  $a$ or $\Delta t$ or $a \Delta t$ is small enough, we can neglect the second term, and 
 approximately define the inertial frame of reference.
 
 Second, Maxwell's equations for the static media system is the traditional 
 Maxwell's equations, and the co-moving media system is in the inertial frame of reference as defined above.
 Therefore, we think that Maxwell's equations for the moving media system are 
 still covariant, {\it i.e.}, are still the traditional Maxwell's equations, which are
 consistent with the two fundamental postulates of special relativity. From the theoretical point
 of view, Maxwell's equations describe the massless photon field, and the basic principles of 
 special relativity should be valid for all the inertial frames. In the next Section, for small velocity,
 we shall prove it explicitly by considering the Lorentz transformation at the order ${\cal O} (v)$,
 and we believe that it should be valid at any order ${\cal O} (v^n)$. 
 
 Third, if we compare the traditional Maxwell's equations with those in Ref~\cite{Wang:2021p2},
 the key difference is that  ${{\partial}\over {\partial t}}$ changing to ${{\partial}\over {\partial t}}+\vec{v}\cdot\nabla$.
 Interestingly, under the Lorentz transformation from the lab frame to the co-moving frame of the media, at the order ${\cal O} (v)$ ($\vec{v}$ is the velocity of media in the lab frame) we have
 \begin{align}
 &\frac{\partial}{\partial t}=\frac{\partial}{\partial t^\prime}+\vec{v}\cdot\nabla^\prime\equiv\frac{d}{dt^\prime}~,\\
 &\vec{v}=\frac{d \vec{x}^\prime}{d t^\prime}~,
 \end{align}
 where $(\vec{x}^\prime,t^\prime)$ and $(\vec{x},t)$ are the times and locations of some point of the media in the lab and co-moving frame, respectively.
 What is more,
 the Faraday's Law of electromagnetic induction for a loop attached to the moving media at the order ${\cal O} (v)$
 gives 
 \begin{align}
\oint \vec{E}(\vec{x},t) \cdot d\vec{l}^\prime &= -\frac{d}{dt^\prime}\int\int \vec{B}^{\prime}(\vec{x}^\prime,t^\prime)\cdot d\vec{S}^\prime\nonumber\\
&=-\int\int \left( \frac{\partial}{\partial t^\prime}+\vec{v}\cdot\nabla^\prime\right) \vec{B}^{\prime}(\vec{x}^\prime,t^\prime)\cdot d\vec{S}^\prime~,
\end{align}   
 where the electric field $\vec{E}$ and the magnetic field $\vec{B}^\prime$ are in the co-moving frame and the lab frame, respectively, and $d\vec{l}^\prime$ and  $d\vec{S}^\prime$ are attached to the moving media in the lab frame~\cite{Jackson:1999}. Therefore,
 we conjecture that the electric and magnetic fields in the expanded Maxwell's equations 
 are not in the same reference frame such as lab or co-moving frame of media. And we shall prove
 this conjecture in the following Sections.
 
 Fourth, from the classical field theory point of view, 
 the Lorentz force is not related to Maxwell's equations directly, and we do not combine it with
 the electric field. Considering a single particle with electric charge $Qe$ and mass $m$ 
 along the worldline $\vec{x}^{\prime }=\vec{x}^{\prime}(t^\prime)$, we can define its four-velocity and four-momentum in the lab frame as 
 \begin{align}
 U^{\mu} \equiv {{dx^{\prime\mu}}\over d\tau}~,~~~
 p^{\mu} \equiv m U^{\mu}~,\label{U}
 \end{align}
 where $\tau$ is the proper time. The relativistic four-force $f^{\mu}$ acting on the particle in the lab frame
 is given by
 \begin{align}
 f^{\mu} = {{dp^{\mu}}\over d\tau} \equiv Qe F^{\prime\mu \nu} U_{\nu}~, 
 \end{align}
 where $F^{\prime\mu \nu}$ is made up of $\vec{E}^\prime$ and $\vec{B}^\prime$ in the lab frame.
 And we get 
 \begin{align}
 \vec{f} = Qe \gamma \left(\vec{E}^\prime(\vec{x}^\prime,t^\prime)+ \vec{v} \times \vec{B}^\prime(\vec{x}^\prime,t^\prime)\right)~,~
 \end{align}
 where $\vec{v}=d\vec{x}^\prime/dt^\prime$. Therefore, we do not need to combine the 
 Lorentz force with the electric field in Maxwell's equations 
 in our approach.
 
Historically, Hertz proposed an expanded Maxwell's equations to satisfy 
Galilean invariance~\cite{Hertz:1890}. Hertz's equations are equivalent to the equations obtained by replacing the partial time derivatives with the total time derivatives wherever the former appear in Maxwell's equations~\cite{Rozov:2015, Rozov:2017}. 
So, Hertz's electrodynamics is a formally Galilean-invariant covering theory of Maxwell's vacuum electrodynamics.
Naively, it seems that Eqs.~\eqref{w1} to \eqref{w4} are the expanded Hertz's equations for moving media. 
However, Hertz's electrodynamics is an invariant theory under the Galileo transformation but not a covariant theory as the modern physics, 
and have been proved to be wrong. Therefore, we shall not study it along this line.

\section{Covariance of Maxwell's Equations under the first-order approximation of media speed}\label{sec4}

In this Section, we will prove that Maxwell's equations are covariant  by considering the Lorentz transformation from the co-moving frame of the media to the lab frame at the order ${\cal O} (v)$.

Assuming the velocity of media is $\vec{v}$ ($v\ll c$) in the lab frame. The relations between times, locations, fields, and operators in the lab and co-moving frames at the order ${\cal O} (v)$ are summarized in Table~\ref{Table_Classical_Results}.
\begin{table}[htb]
	\centering
	\begin{tabular}{c|c}  
		\hline\hline
		 Co-moving frame & Lab frame \\
		\hline 
		 $\vec{x}=\vec{x}^\prime-\vec{v}t^\prime$ & $\vec{x}^\prime=\vec{x}+\vec{v}t$ \\
		 $t=t^\prime-\vec{v}\cdot\vec{x}^\prime$  &  $t^\prime=t+\vec{v}\cdot\vec{x}$  \\
		 $\frac{d\vec{x}}{d t}=0$  &  $\frac{d\vec{x}^\prime}{d t^\prime}=\vec{v}$  \\
		 $\vec{J}_f(\vec{x},t)=\vec{J}_f^\prime(\vec{x}^\prime,t^\prime)-\vec{v}\rho_f^\prime(\vec{x}^\prime,t^\prime)$ & $\vec{J}_f^\prime(\vec{x}^\prime,t^\prime)=\vec{J}_f(\vec{x},t)+\vec{v}\rho_f(\vec{x},t)$ \\
		 $\rho_f(\vec{x},t)=\rho_f^\prime(\vec{x}^\prime,t^\prime)-\vec{v}\cdot\vec{J}_f^\prime(\vec{x}^\prime,t^\prime)$  &  $\rho_f^\prime(\vec{x}^\prime,t^\prime)=\rho_f(\vec{x},t)+\vec{v}\cdot\vec{J}_f(\vec{x},t)$  \\
		 $\vec{E}(\vec{x},t)=\vec{E}^\prime(\vec{x}^\prime,t^\prime)+\vec{v}\times\vec{B}^\prime(\vec{x}^\prime,t^\prime)$ & $\vec{E}^\prime(\vec{x}^\prime,t^\prime)=\vec{E}(\vec{x},t)-\vec{v}\times\vec{B}(\vec{x},t)$ \\
		 $\vec{B}(\vec{x},t)=\vec{B}^\prime(\vec{x}^\prime,t^\prime)-\vec{v}\times\vec{E}^\prime(\vec{x}^\prime,t^\prime)$ & $\vec{B}^\prime(\vec{x}^\prime,t^\prime)=\vec{B}(\vec{x},t)+\vec{v}\times\vec{E}(\vec{x},t)$ \\
		 $\vec{D}(\vec{x},t)=\vec{D}^\prime(\vec{x}^\prime,t^\prime)+\vec{v}\times\vec{H}^\prime(\vec{x}^\prime,t^\prime)$ & $\vec{D}^\prime(\vec{x}^\prime,t^\prime)=\vec{D}(\vec{x},t)-\vec{v}\times\vec{H}(\vec{x},t)$ \\
		 $\vec{H}(\vec{x},t)=\vec{H}^\prime(\vec{x}^\prime,t^\prime)-\vec{v}\times\vec{D}^\prime(\vec{x}^\prime,t^\prime)$ & $\vec{H}^\prime(\vec{x}^\prime,t^\prime)=\vec{H}(\vec{x},t)+\vec{v}\times\vec{D}(\vec{x},t)$ \\
		 $\nabla=\nabla^\prime+\vec{v}\frac{\partial}{\partial t^\prime}$ & $\nabla^\prime=\nabla-\vec{v}\frac{\partial}{\partial t}$ \\
		 $\frac{\partial}{\partial t}=\frac{\partial}{\partial t^\prime}+\vec{v}\cdot\nabla^\prime$  
		 & $\frac{\partial}{\partial t^\prime}=\frac{\partial}{\partial t}-\vec{v}\cdot\nabla$ \\
		  $\frac{d}{d t}=\frac{\partial}{\partial t}$  
		 & $\frac{d}{d t^\prime}=\frac{\partial}{\partial t^\prime}+\vec{v}\cdot\nabla^\prime$ \\
		\hline\hline
	\end{tabular}
	\caption{The known result of low speed ($v\ll c$) approximation of Lorentz transformation at the order ${\cal O} (v)$.}
    \label{Table_Classical_Results}
\end{table}

 Maxwell's equations in co-moving frame of media are Eqs.~(\ref{m1}) to (\ref{m4}). We derive Maxwell's equations in the lab frame at the order ${\cal O} (v)$.
 \begin{align}
 	 \nabla^{\prime} \cdot \vec{D}^{\prime} &=\left(\nabla-\vec{v} \frac{\partial}{\partial t}\right) \cdot(\vec{D}-\vec{v} \times \vec{H})\nonumber \\
 	&=\rho_{f}-\vec{v} \cdot\left(\nabla\times \vec{H}-\vec{J}_{f}\right)+\vec{v} \cdot \nabla\times \vec{H} \nonumber\\
 	&=\rho_{f}^{\prime} \label{mp1}
 \end{align} 
 \begin{align}
 	 \nabla^{\prime} \cdot \vec{{B}^{\prime}} &=\left({\nabla}-\vec{{v}} \frac{{\partial}}{\partial {t}}\right) \cdot(\vec{{B}}+\vec{{v}} \times \vec{{E}}) \nonumber\\
 	&=0+\vec{{v}} \cdot({\nabla} \times \vec{{E}})-\vec{{v}} \cdot({\nabla} \times \vec{{E}})\nonumber\\
 	&=0 \label{mp2}
 \end{align}
 \begin{align}
 	 \nabla^{\prime} \times \vec{E}^{\prime}&=\left(\nabla-\vec{v} \frac{\partial}{\partial t}\right) \times(\vec{E}-\vec{v} \times \vec{B}) \nonumber\\
 	&=-\frac{\partial \vec{B}}{\partial t}-\vec{v} \times\frac{\partial \vec{E}}{\partial t}+(\vec{v} \cdot \nabla) \vec{B}\nonumber\\ &=-\left(\frac{\partial}{\partial t}-\vec{v} \cdot \nabla\right)(\vec{B}+\vec{v} \times \vec{E})\nonumber\\
   &= -\frac{\partial \vec{B}^{\prime}}{\partial t^{\prime}} \label{mp3}
 \end{align}

 \begin{align}
 	\nabla^{\prime} \times \vec{{H}}^{\prime}
 	&=\left({\nabla}-\vec{{v}} \frac{{\partial}}{\partial{t}}\right) \times(\vec{{H}}+\vec{{v}} \times \vec{{D}}) \nonumber\\
 	&=\vec{{J}}_{{f}}+\frac{\partial(\vec{{D}}-\vec{{v}} \times \vec{{H}})}{\partial {t}}+\vec{{v}} {\rho}_{{f}}-(\vec{{v}} \cdot {\nabla}) \vec{{D}}^\prime \nonumber\\
 	&=\vec{{J}}_{{f}}^{\prime}+\frac{\partial \vec{{D}}^{\prime}}{\partial {t}^{\prime}} \label{mp4}
 \end{align}
 We have used the mathematical identity
 \begin{align}
 \nabla\times(\vec{a}\times\vec{b})=(\vec{b}\cdot \nabla)\vec{a}+(\nabla\cdot \vec{b})\vec{a}-(\vec{a}\cdot \nabla)\vec{b}-(\nabla\cdot \vec{a})\vec{b}
 \end{align}
 in the above derivations. 
 
 According to Eqs.~(\ref{mp1}) to (\ref{mp4}), we can conclude that Maxwell's equations in the lab and co-moving frames have the same forms at the order ${\cal O} (v)$. So, we cannot get  Eqs.~(\ref{w3}) and (\ref{w4}) under the Lorentz transformation if all the fields in the equations are from the same reference frame.
 
 \section{Covariance of Maxwell's Equations under the Galileo approximation}\label{sec5}
 
 In this Section, we will prove that Maxwell's equations are covariant under the Galileo transformation.
 
 Assuming the velocity of media is $\vec{v}$ ($v/c\ll 1$) in the lab frame. Under the Galileo transformation, the relations between times, locations, fields, and operators in the lab and co-moving frames are summarized in Table~\ref{Galileo_transformation}, which can be obtained by ignoring the terms proportional to $v/c$ (these terms are apparent by using the SI units, for example, $t^\prime=t+\frac{\vec{v}}{c}\cdot\frac{\vec{x}}{c}\simeq t$, $\vec{D}^\prime=\vec{D}-\frac{\vec{v}}{c}\times\frac{\vec{H}}{c}\simeq \vec{D}$ ) in Table~\ref{Table_Classical_Results}.
 \begin{table}[htb]
 	\centering
 	\begin{tabular}{c|c}  
 		\hline\hline
 		Co-moving frame & Lab frame \\
 		\hline 
 		$\vec{x}=\vec{x}^\prime-\vec{v}t^\prime$ & $\vec{x}^\prime=\vec{x}+\vec{v}t$ \\
 		$t=t^\prime$  &  $t^\prime=t$  \\
 		$\frac{d\vec{x}}{d t}=0$  &  $\frac{d\vec{x}^\prime}{d t^\prime}=\vec{v}$  \\
 		$\vec{J}_f(\vec{x},t)=\vec{J}_f^\prime(\vec{x}^\prime,t^\prime)-\vec{v}\rho_f^\prime(\vec{x}^\prime,t^\prime)$ & $\vec{J}_f^\prime(\vec{x}^\prime,t^\prime)=\vec{J}_f(\vec{x},t)+\vec{v}\rho_f(\vec{x},t)$ \\
 		$\rho_f(\vec{x},t)=\rho_f^\prime(\vec{x}^\prime,t^\prime)$  &  $\rho_f^\prime(\vec{x}^\prime,t^\prime)=\rho_f(\vec{x},t)$  \\
 		$\vec{E}(\vec{x},t)=\vec{E}^\prime(\vec{x}^\prime,t^\prime)+\vec{v}\times\vec{B}^\prime(\vec{x}^\prime,t^\prime)$ & $\vec{E}^\prime(\vec{x}^\prime,t^\prime)=\vec{E}(\vec{x},t)-\vec{v}\times\vec{B}(\vec{x},t)$ \\
 		$\vec{B}(\vec{x},t)=\vec{B}^\prime(\vec{x}^\prime,t^\prime)$ & $\vec{B}^\prime(\vec{x}^\prime,t^\prime)=\vec{B}(\vec{x},t)$ \\
 		$\vec{D}(\vec{x},t)=\vec{D}^\prime(\vec{x}^\prime,t^\prime)$ & $\vec{D}^\prime(\vec{x}^\prime,t^\prime)=\vec{D}(\vec{x},t)$ \\
 		$\vec{H}(\vec{x},t)=\vec{H}^\prime(\vec{x}^\prime,t^\prime)-\vec{v}\times\vec{D}^\prime(\vec{x}^\prime,t^\prime)$ & $\vec{H}^\prime(\vec{x}^\prime,t^\prime)=\vec{H}(\vec{x},t)+\vec{v}\times\vec{D}(\vec{x},t)$ \\
 		$\nabla=\nabla^\prime$ & $\nabla^\prime=\nabla$ \\
 		$\frac{\partial}{\partial t}=\frac{\partial}{\partial t^\prime}+\vec{v}\cdot\nabla^\prime$  
 		& $\frac{\partial}{\partial t^\prime}=\frac{\partial}{\partial t}-\vec{v}\cdot\nabla$ \\
 		$\frac{d}{d t}=\frac{\partial}{\partial t}$  
 		& $\frac{d}{d t^\prime}=\frac{\partial}{\partial t^\prime}+\vec{v}\cdot\nabla^\prime$ \\
 		\hline\hline
 	\end{tabular}
 	\caption{Relations under the Galileo transformation between the co-moving and  lab frames.}
 	\label{Galileo_transformation}
 \end{table}
 
 Maxwell's equations in co-moving frame of media are Eqs.~(\ref{m1}) to (\ref{m4}). We derive Maxwell's equations in the lab frame under the Galileo transformation.
 \begin{align}
 \nabla^{\prime} \cdot \vec{D}^{\prime} &=\nabla\cdot\vec{D}=\rho_{f}=\rho_{f}^{\prime} \label{G1}
 \end{align} 
 \begin{align}
 \nabla^{\prime} \cdot \vec{{B}^{\prime}} &=\nabla\cdot \vec{B}=0 \label{G2}
 \end{align}
 \begin{align}
 \nabla^{\prime} \times \vec{E}^{\prime}&=\nabla\times(\vec{E}-\vec{v} \times \vec{B}) \nonumber\\
 &=-\left(\frac{\partial}{\partial t}-\vec{v} \cdot \nabla\right)\vec{B}\nonumber\\
 &= -\frac{\partial \vec{B}^{\prime}}{\partial t^{\prime}} \label{G3}
 \end{align}
 
 \begin{align}
 \nabla^{\prime} \times \vec{{H}}^{\prime}
 &={\nabla}\times(\vec{{H}}+\vec{{v}} \times \vec{{D}}) \nonumber\\
 &=\vec{{J}}_{{f}}+\frac{\partial\vec{{D}}}{\partial {t}}+\vec{{v}} {\rho}_{{f}}-(\vec{{v}} \cdot {\nabla}) \vec{{D}}^\prime \nonumber\\
 &=\vec{{J}}_{{f}}^{\prime}+\frac{\partial \vec{{D}}^{\prime}}{\partial {t}^{\prime}} \label{G4}
 \end{align}
 
 Eqs.~(\ref{G1}) to (\ref{G4}) show that Maxwell's equations are covariant under the Galileo transformation from the co-moving frame to the lab frame, which means we cannot get  Eqs.~(\ref{w3}) and (\ref{w4}) under the Galileo transformation if all the fields in the equations are from the same reference frame.

\section{Different reference frames viewpoint under the first-order approximation of media speed}\label{sec6}
 
The previous two Sections show that neither the Lorentz transformation nor the Galileo transformation can lead to Eqs.~(\ref{w3}) and (\ref{w4}) if all the fields in the equations are from the same reference frame.
In Sec.~\ref{sec3}, we conjecture that the electric and magnetic fields in the expanded Maxwell's equations 
 are not in the same reference frame such as lab or co-moving frame of media. And we shall prove
 this conjecture in the following study.
To be concrete, we will show that the equations similar with but different from Eqs.~\eqref{w3} and \eqref{w4} 
can be obtained under Lorentz transformation at the order ${\cal O} (v)$ from the sense of different reference frames. 

Based on  Eqs.~(\ref{m1}) to (\ref{m4}) and Eqs.~(\ref{mp1}) to (\ref{mp4}), at the order ${\cal O} (v)$ we can get another two equations
 \begin{align}
\nabla^\prime\times\vec{E}(\vec{x},t)&=\nabla ^\prime \times \left( \vec{E}^\prime(\vec{x}^\prime,t^\prime)+\vec{v}\times\vec{B}^\prime(\vec{x}^\prime,t^\prime) \right)\nonumber\\
&=-\left( \frac{\partial }{\partial t^\prime}+\vec{v}\cdot\nabla^\prime\right) \vec{B}^{\prime}(\vec{x}^\prime,t^\prime)\label{wm3}\\
&=- \frac{d }{d t^\prime} \vec{B}^{\prime}(\vec{x}^\prime,t^\prime)~.\label{wmp3}
\end{align}  
 \begin{align}
\nabla^\prime \times\vec{{H}}(\vec{x},t) &= \nabla^\prime \times \left( \vec{H}^\prime(\vec{x}^\prime,t^\prime)-\vec{v}\times\vec{D}^\prime(\vec{x}^\prime,t^\prime) \right)\nonumber\\
&=\vec{{J}}_{{f}}^{\prime}(\vec{x}^\prime,t^\prime) + \frac{\partial }{\partial t^\prime}\vec{D}^\prime(\vec{x}^\prime,t^\prime)  - \vec{{v}}\rho^\prime(\vec{x}^\prime,t^\prime)+ (\vec{v}\cdot\nabla^\prime) \vec{D}^{\prime}(\vec{x}^\prime,t^\prime) \nonumber\\
&=\vec{{J}}_{{f}}(\vec{x},t) +\left( \frac{\partial }{\partial t^\prime}+\vec{v}\cdot\nabla^\prime\right) \vec{D}^{\prime}(\vec{x}^\prime,t^\prime)~\label{wm4}\\
&=\vec{{J}}_{{f}}(\vec{x},t) + \frac{d }{d t^\prime} \vec{D}^{\prime}(\vec{x}^\prime,t^\prime)~.\label{wmp4}
\end{align} 

Though Eqs.~(\ref{wm3}) and (\ref{wm4}) as well as Eqs.~(\ref{mp1}) and (\ref{mp2}), summarized as
 \begin{align}
  &\nabla^{\prime} \cdot \vec{D}^{\prime} (\vec{x}^\prime,t^\prime)=\rho_{f}^{\prime}(\vec{x}^\prime,t^\prime)~,\\
  &\nabla^{\prime} \cdot \vec{B}^{\prime} (\vec{x}^\prime,t^\prime)=0~,\\
 &\nabla^\prime\times\vec{E}(\vec{x},t)=-\left( \frac{\partial }{\partial t^\prime}+\vec{v}\cdot\nabla^\prime\right) \vec{B}^{\prime}(\vec{x}^\prime,t^\prime)~,\label{vv1}\\
&\nabla^\prime \times\vec{{H}}(\vec{x},t) =\vec{{J}}_{{f}}(\vec{x},t) +\left( \frac{\partial }{\partial t^\prime}+\vec{v}\cdot\nabla^\prime\right) \vec{D}^{\prime}(\vec{x}^\prime,t^\prime)~,\label{vv2}
\end{align} 
are similar to Eqs.~(\ref{w1}) to (\ref{w4}) in forms, we have to emphasize that $\rho_{f}^\prime(\vec{x}^\prime,t^\prime)$, $\vec{D}^{\prime}(\vec{x}^\prime,t^\prime)$, $\vec{B}^{\prime}(\vec{x}^\prime,t^\prime)$, $\nabla^\prime$, and $\frac{\partial}{\partial t^\prime}$ are in the lab frame, and $\vec{J}_f(\vec{x},t)$, $\vec{E}(\vec{x},t)$, and $\vec{H}(\vec{x},t)$ are in the co-moving frame of media for our equations. This might be different from the viewpoint of Ref~\cite{Wang:2021p2}, where all the fields and operators seem to be regarded as in the same frame by default.
Interestingly, 
note that $\vec{E}(\vec{x},t)=\vec{E}^\prime(\vec{x}^\prime,t^\prime)+\vec{v}\times\vec{B}^\prime(\vec{x}^\prime,t^\prime)$,
we find that the electric force in co-moving frame is the combination of the electric force and Lorentz force 
in the lab frame, and then our equations might be useful in the application.

Based on Eqs.~(\ref{wmp3}) and (\ref{wmp4}), we, at the order ${\cal O} (v)$, get the Faraday's Law of electromagnetic induction and Ampere-Maxwell law for the moving media in the lab frame
 \begin{align}
&\oint \vec{E}(\vec{x},t) \cdot d\vec{l}^\prime = -\frac{d}{dt^\prime}\int\int \vec{B}^{\prime}(\vec{x}^\prime,t^\prime)\cdot d\vec{S}^\prime~,\\
&\oint \vec{H}(\vec{x},t) \cdot d\vec{l}^\prime = \int\int\vec{J}_f(\vec{x},t) \cdot d\vec{S}^\prime+\frac{d}{dt^\prime}\int\int \vec{D}^{\prime}(\vec{x}^\prime,t^\prime)\cdot d\vec{S}^\prime~,
\end{align}  
where  $d\vec{l}^\prime$ and  $d\vec{S}^\prime$ are attached to the moving media in the lab frame.  

Eqs.~(\ref{vv1}) and (\ref{vv2}) are approximate results at the order ${\cal O} (v)$. According to the derivation in Appendix~\ref{app}, the precise equations involving fields in different reference frames are
\begin{align}
\nabla^{\prime} \times\left(\vec{E}-(\gamma-1) \vec{E}_{\perp}^{\prime}\right)&=-\left(\frac{\partial}{\partial t^{\prime}}+\gamma \vec{v} \cdot \nabla^{\prime}\right) \vec{B}^{\prime}~, \label{yy1}\\
\nabla^{\prime} \times\left(\vec{H}-(\gamma-1) \vec{H}_{\perp}^{\prime}\right)&=\vec{J}_{f}-(\gamma-1) \vec{J}_{f_{\|}}^{\prime}+\left(\frac{\partial}{\partial t^{\prime}}+\gamma \vec{v} \cdot \nabla^{\prime}\right) \vec{D}^{\prime}~.\label{yy2}
\end{align}
We can take $\gamma=1$ in Eqs.~(\ref{yy1}) and (\ref{yy2}) at the order ${\cal O} (v)$, which directly leads to Eqs.~(\ref{vv1}) and (\ref{vv2}).

\section{Another variant of Maxwell's equations with some redefined fields}\label{sec7}

 In this Section, we will explain another variant of Maxwell's equations at a low speed with some redefined fields.
 
 Using $F^\prime_{\mu\nu}$, $K^\prime_{\mu\nu}$, $U^\mu$ ($F^\prime_{\mu\nu}$ and $K^\prime_{\mu\nu}$ are made up of fields in the lab frame, and $U^\mu$ is defined in Eq.~(\ref{U})), we can construct new four-vector $F^\prime_{\mu\nu}U^\mu$ and $K^\prime_{\mu\nu}U^\mu$, and get the extension of the relation $\vec{D}=\epsilon \vec{E}$ for the moving media in the lab frame~\cite{Landau-Lifshitz:ED_for_CM:SC}
  \begin{align}
 	 K^\prime_{\mu\nu}U^\mu=\epsilon F^\prime_{\mu\nu}U^\mu~.
 \end{align}
 Similarly, the extension of the relation $\vec{B}=\mu \vec{H}$ for the moving media in the lab frame~\cite{Landau-Lifshitz:ED_for_CM:SC} is
\begin{align}
	F^\prime_{\kappa\nu}U_\lambda + F^\prime_{\nu\lambda}U_\kappa + F^\prime_{\lambda\kappa}U_\nu= \mu \left(K^\prime_{\kappa\nu}U_\lambda + K^\prime_{\nu\lambda}U_\kappa + K^\prime_{\lambda\kappa}U_\nu\right)~.  
\end{align}
Reduced to three-vector form, we can get
 \begin{align}
 	&\vec{D}^\prime(\vec{x}^\prime,t^\prime) + \vec{v} \times \vec{H}^\prime(\vec{x}^\prime,t^\prime) = \epsilon \left( \vec{E}^\prime + \vec{v}  \times \vec{B}^\prime(\vec{x}^\prime,t^\prime) \right)~, \label{new1}\\
 	&\vec{B}^\prime(\vec{x}^\prime,t^\prime) - \vec{v} \times \vec{E}^\prime(\vec{x}^\prime,t^\prime) = \mu \left( \vec{H}^\prime(\vec{x}^\prime,t^\prime) - \vec{v}  \times \vec{D}^\prime(\vec{x}^\prime,t^\prime) \right)~, 	\label{new2}
 \end{align}
 where $\vec{v}=\frac{d\vec{x}^\prime}{dt^\prime}$.
 Eqs.~(\ref{new1}) and (\ref{new2})  are the new constitutive equations of the moving media in the lab frame. And we emphasize that $\epsilon$ and $\mu$ are the electric permittivity and magnetic permeability of the media at the co-moving frame, respectively. 
 
 When $v\ll c$, at the order ${\cal O} (v)$ we get
  \begin{align}
 	\vec{D}^\prime(\vec{x}^\prime,t^\prime) 
 	&=\epsilon \left( \vec{E}^\prime(\vec{x}^\prime,t^\prime) + \alpha \vec{v}  \times\vec{B}^\prime(\vec{x}^\prime,t^\prime) \right)~,\\
	\vec{B}^\prime (\vec{x}^\prime,t^\prime) &=\mu \left( \vec{H}^\prime(\vec{x}^\prime,t^\prime) - \alpha\vec{v} \times \vec{D}^\prime(\vec{x}^\prime,t^\prime) \right)~,
 \end{align}
 with
  \begin{align}
 \alpha ={{\epsilon_r \mu_r -1} \over \epsilon_r \mu_r}~.
 \end{align}
 It is noted that $\epsilon_r=\epsilon$ and $\mu_r=\mu$ when the Natural Units are used.
 If we define 
 \begin{align}
 	&\vec{E}^\star=\vec{D}^\prime/  \epsilon~,   \\
 	&\vec{H}^\star=\vec{B}^\prime / \mu~, 
 \end{align}
 it is direct to get
  \begin{align}
 	&\vec{E}^\star = \vec{E}^\prime + \alpha \vec{v} \times \vec{B}^\prime~, \\
 	&\vec{H}^\star = \vec{H}^\prime - \alpha \vec{v} \times \vec{D}^\prime~,
 \end{align}
 at the order ${\cal O} (v)$.
 Then, we have
    \begin{align}
  	&\nabla^\prime \times \vec{E}^\star = - \frac{\partial \vec{B^\prime}}{\partial t^\prime} + \nabla^\prime \times  \left( \alpha \vec{v} \times \vec{B}^\prime \right)~,\label{r1} \\
  	&\nabla^\prime \times \vec{H}^\star = \vec{J}^\prime_f+\frac{\partial \vec{D^\prime}}{\partial t^\prime} - \nabla^\prime \times \left (\alpha \vec{v} \times \vec{D}^\prime\right) ~.\label{r2} 
  \end{align}
  
  The results of Eqs.~(\ref{r1}) and (\ref{r2}) are totally similar to those 
in Refs~\cite{Rozov:2015, Rozov:2017} in forms~\footnote{Please check the formula (14) and (15) in Ref~\cite{Rozov:2015}, or formula (9) and (13) in Ref~\cite{Rozov:2017}. }. But, as demonstrated by us, these formulas cannot be obtained in the standard framework of classical electrodynamics if all the fields are normally defined. 
Although the contents in this Section have no direct relation with the Ref~\cite{Wang:2021p2}, 
we want to remind the readers that, given some special definitions, we can easily get some new variants 
of the standard Maxwell's equations. Even if the new variants might be useful and meaningful at some special applied scenarios, they are subtle and misleading, sometimes are not even self-consistent. 
In short,  we should treat them very carefully.

\section{Conclusion}\label{sec8}

The classical electrodynamics, special relativity, and Lorentz covariance requirements are 
the big science revolutions created by Maxwell, Einstein, and many other scientific figures, 
which have stood the test of time. 

From the above theoretical derivations and analysis, we showed that people cannot get the results in Ref~\cite{Wang:2021p2} in the standard framework of Maxwell's theory under the ${\cal O} (v)$ approximation or under the Galileo approximation, if all the relevant physical quantities are defined in the same reference frame.   

We have shown that equations similar to those of Ref~\cite{Wang:2021p2} in forms can be obtained from the viewpoint of different reference frames, {\it i.e.}, 
the physical quantities in our equations are defined in different frames. Therefore, our understanding might be not consistent with the viewpoint of Ref~\cite{Wang:2021p2}, where all the fields and operators seem to be defined in the same reference frame by default.

We hope our present work enhances the mutual understanding between theoretical physics community and material science community. And we wish that our timely and appropriate (and maybe inappreciable) clarification will promote the reasonable application of classical electrodynamics in material and/or other applied sciences.

\appendix
\section{Derivation for Eqs.~(\ref{yy1}) and (\ref{yy2})}\label{app} 

Under the Lorentz transformation from the lab frame to the co-moving frame, the fields transform as
\begin{align}
	&\vec{E}=\gamma\left(\vec{E}^{\prime}+\vec{v} \times \vec{B}^{\prime}\right)-(\gamma-1) \vec{E}_{\|}^{\prime}~, \\
	&\vec{D}=\gamma\left(\vec{D}^{\prime}+\vec{v} \times \vec{H}^{\prime}\right)-(\gamma-1) \vec{D}_{\|}^{\prime}~, \\
	&\vec{B}=\gamma\left(\vec{B}^{\prime}-\vec{v} \times \vec{E}^{\prime}\right)-(\gamma-1) \vec{B}_{\|}^{\prime}~, \\
	&\vec{H}=\gamma\left(\vec{H}^{\prime}-\vec{v} \times \vec{D}^{\prime}\right)-(\gamma-1) \vec{H}_{\|}^{\prime}~,
\end{align}
where $\gamma=\frac{1}{\sqrt{1-v^{2}}}$. And for the operators, we have 
\begin{align}
	\nabla&=\nabla^{\prime}+\frac{\gamma^{2}}{\gamma+1}\left(\vec{v} \cdot \nabla^{\prime}\right) \vec{v}+\gamma \vec{v} \frac{\partial}{\partial t^{\prime}}~, \\
	\frac{\partial}{\partial t}&=\gamma\left(\frac{\partial}{\partial t^{\prime}}+\vec{v} \cdot \nabla^{\prime}\right)~.
\end{align}
Then, it is direct to get
\begin{align}
	\nabla \times \vec{E}=&\nabla^{\prime} \times \vec{E}-\gamma(\gamma-1)\left(\frac{\partial}{\partial t^{\prime}}+\vec{v} \cdot \nabla^{\prime}\right) \vec{B}_{\perp}^{\prime} \nonumber\\
	&+\left(\vec{v} \cdot \nabla^{\prime}\right)\left(\frac{\gamma^{3}}{\gamma+1} \vec{v} \times \vec{E}^{\prime}\right)+\frac{\partial}{\partial t^{\prime}}\left(\gamma^{2} \vec{v} \times \vec{E}^{\prime}-(\gamma-1) \vec{B}_{\perp}^{\prime}\right)~,\\
	-\frac{\partial \vec{B}}{\partial t}=&-\gamma^{2}\left(\frac{\partial}{\partial t^{\prime}}+\vec{v} \cdot \nabla^{\prime}\right) \vec{B}^{\prime}+\gamma(\gamma-1)\left(\frac{\partial}{\partial t^{\prime}}+\vec{v} \cdot \nabla^{\prime}\right) \vec{B}_{\|}^{\prime} \nonumber\\
	&+\frac{\partial}{\partial t^{\prime}}\left(\gamma^{2} \vec{v} \times \vec{E}^{\prime}\right)+\left(\vec{v} \cdot \nabla^{\prime}\right)\left(\gamma^{2} \vec{v} \times \vec{E}^{\prime}\right)~,
\end{align}
where we have used
\begin{align}
	\vec{v} \times \vec{a}_{\|} &=0~, \\
	\vec{v} \times(\vec{v} \times \vec{a}) &=-v^{2} \vec{a}_{\perp}~.
\end{align}
By using Eq.~(\ref{m3}) and
\begin{align}
	\left(\vec{v} \cdot \nabla^{\prime}\right)\left(\vec{v} \times \vec{E}^{\prime}\right)&=-\nabla^{\prime}\times\left(\vec{v}\times \left(\vec{v} \times \vec{E}^{\prime}\right)\right) -\vec{v}\cdot\left(\nabla^\prime \times \vec{E}^{\prime} \right) \vec{v}\nonumber\\
	&=\frac{\gamma^2-1}{\gamma^2}\left(\nabla^{\prime}\times \vec{E}^{\prime}_\perp +\frac{\partial}{\partial t^{\prime}} \vec{B}_{\|}^\prime \right) ~,
\end{align}
we get
\begin{equation}
	\nabla^{\prime} \times\left(\vec{E}-(\gamma-1) \vec{E}_{\perp}^{\prime}\right)=-\left(\frac{\partial}{\partial t^{\prime}}+\gamma \vec{v} \cdot \nabla^{\prime}\right) \vec{B}^{\prime}~. \label{y1}
\end{equation}
Taking a similar approach as that to get Eq.~(\ref{y1}), we can also obtain
\begin{align}
	\nabla^{\prime} \times\left(\vec{H}-(\gamma-1) \vec{H}_{\perp}^{\prime}\right)&=\vec{J}_{f}-(\gamma-1) \vec{J}_{f_{\|}}^{\prime}+\left(\frac{\partial}{\partial t^{\prime}}+\gamma \vec{v} \cdot \nabla^{\prime}\right) \vec{D}^{\prime}~.\label{y2}
\end{align}
So, Eqs.~(\ref{y1}) and (\ref{y2}) are the precise equations involving fields in different reference frames.

\newpage
\section*{Acknowledgement}

This research was supported by Scientific Research Starting Project YJ202011 for Advanced Imported Talents of Wuyi University, by the Projects 11875062, 11947302 and 11875148 supported by the National Natural Science Foundation of China, and by the Key Research Program of Frontier Science, CAS.

%%%%%%%%%%%%%%%%%%%%%

%
%

\end{document}